\documentclass[12pt,twoside]{article}

\usepackage{amssymb}
\usepackage{amsmath}
\usepackage{wrapfig}
\usepackage{graphicx}
\usepackage[eqsecnum]{cmp2e}

\newcommand{\Tr}{\operatorname{Tr}}

\title[Ergodicity in Strongly Correlated Systems]%
{Ergodicity in \protect\\ Strongly Correlated Systems}

\author{A. Avella, F. Mancini and E. Plekhanov}
\address{Dipartimento di Fisica ``E.R. Caianiello'' - Unit\`{a}
CNISM di Salerno \\
Universit\`{a} degli Studi di Salerno, I-84081 Baronissi (SA),
Italy}

\begin{document}

\maketitle

\begin{abstract}
We present a concise, but systematic, review of the ergodicity issue
in strongly correlated systems. After giving a brief historical
overview, we analyze the issue within the Green's function formalism
by means of the equations of motion approach. By means of this
analysis, we are able to individuate the primary source of
non-ergodic dynamics for a generic operator and also to give a
recipe to compute unknown quantities characterizing such a behavior
within the Composite Operator Method. Finally, we present examples
of non-trivial strongly correlated systems where it is possible to
find a non-ergodic behavior.
\pacs 05.30.Jp, 75.10.Jm, 71.10.-w
\end{abstract}

\section{Historical Overview}

The issue of ergodicity in condensed matter physics is well known
since fifties \cite{Kubo_57}. Given two operators $A$ and $B$,
describing physical quantities (e.g., charge, spin, pair densities
or currents), one can study the physical response of a system
described by a certain Hamiltonian $H$ through the generalized
susceptibility
\begin{equation}\label{susceptibility}
\chi_{AB}=\lim_{h \to 0}\frac{\partial \langle A \rangle}{\partial
h}
\end{equation}
where $h$ is an external field entering the Hamiltonian of the
system under study in a coupling term of the type $-hB$ and $\langle
\cdots \rangle$ stands for the statistical average in some ensemble
over the perturbed system. Kubo \cite{Kubo_57} immediately noticed
that the static isolated susceptibility $\chi^I(0)$, defined for an
isolated system perturbed by an external field turned on
adiabatically, and the isothermal susceptibility $\chi^T$, defined
for a system in thermal equilibrium in presence of a
time-independent external field, can be generally different. In
particular, Falk \cite{Falk_68} has shown that the static isolated
susceptibility is just a lower bound for the isothermal one
$\chi^I(0) \leq \chi^T$.

We recall that the static isolated susceptibility $\chi^I(0)$,
within the linear response theory, is defined through the related
retarded Green's function \cite{Kubo_57}
\begin{equation}\label{isolated}
\chi^I(\omega)=-\mathcal{F}\left[\mathrm{i}\,\theta(t_i-t_j)\left\langle
\left[A(t_i),B(t_j)\right]\right\rangle_0\right]
\end{equation}
where $\langle \cdots \rangle_0$ stands for the statistical average
in the microcanonical ensemble (i.e., fixing energy) on the
unperturbed system and $\mathcal{F}$ for the Fourier transform.

On the other hand, the isothermal susceptibility $\chi^T$ can be
computed as
\begin{equation}\label{isothermal}
\chi^T = \int_0^\beta \langle A(-\mathrm{i}\lambda) B \rangle_0 \,
d\lambda - \beta \langle A \rangle_0 \langle B \rangle_0
\end{equation}
In fact, starting from the expression of the thermal average in the
canonical ensemble (i.e., fixing temperature) $\langle A \rangle$
\begin{equation}\label{average}
\langle A \rangle = \frac1Z \Tr\left(A e^{-\beta (H-hB)}\right)
\end{equation}
where $\beta=\frac1T$ and $Z=\Tr\left(e^{-\beta H-hB}\right)$, we
can expand $e^{-\beta (H-hB)}$ in powers of $h$ and get
\begin{equation}
e^{-\beta (H-hB)} \approxeq e^{-\beta H}\left(1 + h \int_0^{\beta}
d\lambda e^{\lambda H} B e^{-\lambda H} + O(h^2)\right)
\end{equation}
Substituting this expansion into (\ref{average}) and retaining only
the first order term in $h$, we get at the numerator
\begin{equation}
\Tr\left(A e^{-\beta (H-hB)}\right) \approxeq Z_0 \langle A
\rangle_0 + h Z_0 \int_0^{\beta} d\lambda \langle A B(i\lambda)
\rangle_0
\end{equation}
and at the denominator
\begin{equation}
Z \approxeq Z_0 \left( 1 + h \beta \langle B \rangle_0 \right)
\end{equation}
and by the ratio
\begin{equation}
\langle A \rangle \approxeq \langle A \rangle_0 + h \int_0^{\beta}
d\lambda \langle A B(i\lambda) \rangle_0 - h\beta \langle A
\rangle_0 \langle B \rangle_0
\end{equation}
where $Z_0$ denotes the partition function of the unperturbed
system. Taking the derivative after (\ref{susceptibility}) and
exploiting the cyclic property of the trace, we obtain the
isothermal susceptibility as in (\ref{isothermal}).

Now, if we rewrite both expressions by means of the general formulas
for the retarded Green's functions and the correlation functions
given in the companion article \cite{Mancini_06} (see Section 3),
present in this same issue, we get
\begin{equation}
\chi^I(0)=\frac{1}{V} \sum_{\mathbf{k},l}
\frac{\sigma^{(l,-1)}(\mathbf{k})}{\omega_l(\mathbf{k})}
\end{equation}
and
\begin{equation}
\chi^T= \beta \frac{1}{V} \sum_{\mathbf{k}} \Gamma_{AB}(\mathbf{k})
+ \frac{1}{V} \sum_{\mathbf{k},l}
\frac{\sigma^{(l,-1)}(\mathbf{k})}{\omega_l(\mathbf{k})} - \beta
\langle A \rangle_0 \langle B \rangle_0
\end{equation}
where \cite{Mancini_06,Mancini_04} $V$ is proportional to the volume
of the system, the sum over $l$ ranges over the number of fields in
the chosen basis, $\sigma^{(l)}$ are the spectral density functions,
$\omega^{(l)}$ are the poles of the propagator, $\Gamma_{AB}$ is an
unknown function appearing in case of poles with zero value.

We can immediately see that the two susceptibilities differ for the
following expression
\begin{equation}\label{diff}
\chi^T-\chi^I(0)= \beta \frac{1}{V} \sum_{\mathbf{k}}
\Gamma_{AB}(\mathbf{k}) - \beta \langle A \rangle_0 \langle B
\rangle_0
\end{equation}

Now, one can check that rewriting the expression $\lim_{t \to
\infty} \langle A B(t) \rangle$ by means of the general formula for
the correlation functions given in the companion article
\cite{Mancini_06} (see Section 3), present in this same issue, we
just get
\begin{equation}
\lim_{t \to \infty} \langle A B(t) \rangle = \frac{1}{V}
\sum_{\mathbf{k}} \Gamma_{AB}(\mathbf{k})
\end{equation}
and, accordingly, the difference at the r.h.s. of (\ref{diff}) is
just what enters Khintchin's theorem \cite{Khintchin_49}: a dynamics
is ergodic (i.e., phase space equilibrium averages are equal to
ensemble microcanical averages, which are much easier to compute)
\begin{equation}
\langle A B \rangle = \int_0^\infty dt \overline{AB(t)}
\end{equation}
if an only if
\begin{equation}\label{erg_def}
\lim_{t \to \infty} \langle A B(t) \rangle = \langle A \rangle
\langle B \rangle
\end{equation}
In other words a dynamic is ergodic if correlations attenuate in
time. In particular, for $B \equiv A$, the dynamics of $A$ is
ergodic if, during its time evolution, it has non-zero matrix
elements only between states within a zero-volume region of the
phase space of the system \cite{Suzuki_71}.

It is clear now the link between ergodicity and response theory: the
two definitions of susceptibility differ when the dynamics of the
system is not ergodic. Two little, but important, notices: finite
systems are not ergodic by definition, just because of the
inequivalence of the ensembles; non-ergodicity at zero temperature
is just the result of a degeneracy in the ground state.

Several years later it was shown \cite{Morita_69,Kwork_69} that the
difference between the two definitions of susceptibility is related
to the zero-frequency anomaly exhibited by bosonic correlation
functions: the presence of undetermined constants in the bosonic
correlation functions. This is exactly what the relations derived
above and the results of the companion article \cite{Mancini_06}
predict establishing a definite link between the ergodicity of the
dynamics and the Green's function formalism. It was first put in
evidence in \cite{Stevens_65} and then studied by many other authors
\cite{Callen_67,Fernandez_67,Morita_69,Kwork_69,Suzuki_71,Ramos_71,Huber_77,Aksenov_78,Aksenov_78a,Aksenov_87}.
There is a general believe that this problem is of academic interest
and in the last years no much attention has been dedicated to it.
The main reason is that the response functions, the experimentally
observed quantities, are given by retarded bosonic Green's function
which formally do not depend on the such undetermined constants,
which are, therefore, considered of no physical interest. The
general attitude \cite{Kubo_57,Callen_67} is to believe that in
macroscopic real systems at equilibrium at a temperature $T$, the
fluctuations are very small and the interaction between the system
and the reservoir would introduce an irreversible relaxation and
decouple the correlation functions. Then, as suggested in
\cite{Callen_67}, these constants should be always determined by
requiring the ergodicity. This procedure is some how an artifice and
may lead to serious problems because it might break the internal
self-consistency of the entire formulation. As remarked in
\cite{Callen_67}, the zero-frequency anomaly is a manifestation of
the difficulty in extracting irreversible behavior from the
statistical mechanics. This is true, but as long as we use the
scheme of statistical mechanics we must be careful in doing
self-consistent calculations. Breaking the self-consistency might
bring to serious errors.

According to the well-known relations existing between casual ($C$),
retarded ($R$) Green's functions and correlation functions
\begin{align}
&\Re [G^{R} ({\bf k}, \omega)]=\Re [G^{C} ({\bf k},
\omega)] \\
&\Im [G^{R} ({\bf k}, \omega)]= \tanh \left ( {{{\beta
\omega } \over 2}} \right)\Im [G^{C} ({\bf k}, \omega)] \\
&C({\bf k}, \omega)=-\left[ {1+\tanh \left ( {{{\beta \omega } \over
2}} \right)} \right]\Im [G^{C} ({\bf k}, \omega)]
\end{align}
the zero-frequency excitations do not contribute explicitly to the
imaginary part of the retarded Green's functions and, consequently,
$\Gamma$ does not explicitly appear in the expressions of
susceptibilities. At any rate, susceptibilities retain an implicit
dependence on $\Gamma$ through the matrix elements. Then, the right
procedure to compute both correlation functions and susceptibilities
is clearly the one that starts from the causal Green's function,
which is the only Green's function that explicitly depends on
$\Gamma$. It is worth noticing that the value of $\Gamma$
dramatically affects the values of directly measurable quantities
(e.g., compressibility, specific heat, magnetic susceptibility,
\ldots) through the values of correlation functions and
susceptibilities. According to this, whenever it is possible,
$\Gamma$ should be exactly calculated case by case.

If we do not have access to the complete set of eigenstates and
eigenvalues of the system, which is the rule in the most interesting
cases, we have to compute correlation functions and susceptibilities
within some, often approximated, analytical framework. Now, since no
analytical tool can easily determine $\Gamma$ (e.g., the equations
of motion cannot be used to fix $\Gamma$ as it is constant in time),
one usually assumes the ergodicity of the dynamics of $\psi$ and
simply substitutes $\Gamma$ by its ergodic value (i.e., by the
r.h.s. of (\ref{erg_def})):
\begin{equation}
\Gamma^{erg}({\bf i},{\bf j})=\langle \psi({\bf i}) \rangle \langle
\psi^{\dagger}({\bf j}) \rangle.
\end{equation}
Unfortunately, this procedure cannot be justified a priori (i.e.,
without computing $\Gamma$ through its definition (\ref{Gammadef}))
by absolutely no means. The existence of just one integral of motion
and, more generally, of any operator that has a diagonal part with
respect to the Hamiltonian \cite{Suzuki_71} (i.e. by any operator
that has a diagonal entries whenever written in the basis of
eigenstates of the Hamiltonian) divides the phase space into
separate subspaces not connected by the dynamics. This latter, in
turn, becomes non ergodic: time averages give different results with
respect to ensemble averages. This latter consideration also
clarifies why the ergodic nature of the dynamics of an operator
mainly depends on the Hamiltonian it is subject to.

It is really remarkable that $\Gamma$ is directly related to
relevant measurable quantities such as compressibility and specific
heat trough the dissipation-fluctuation theorem. For instance, we
recall the formula that relates the compressibility to the total
particle number fluctuations
\begin{equation}  \label{Eq1.36a}
\kappa=\beta \frac{V}{N^2} \left[\langle \hat{N}^2\rangle-N^2\right]
\end{equation}
where $\hat{N}$ is the total particle number operator, $N$ is its
average and $V$ is proportional to the volume of the system. We see
that a compressibility different from zero requires the
non-ergodicity of the system with respect to total particle number
operator. According to this, in the case of infinite systems too the
correct determination of $\Gamma$ cannot be considered as an
irrelevant issue (e.g., (\ref{Eq1.36a}) holds in the thermodynamic
limit too).

In the next section, we provide some examples of violation of the
ergodic condition (\ref{erg_def}). It is necessary pointing out, in
order to avoid any possible confusion to the reader, that we are
using \emph{full} operators and not \emph{fluctuation} ones (i.e.,
we use operators not diminished of their average value, in contrast
with what it is usually done for the bosonic excitations like spin,
charge and pair). According to this, the $\Gamma$ can be different
from zero (i.e., be equal to the squared average of the operator),
and still indicate an ergodic dynamics for the operator.

\section{Examples}

\subsection{Two-site Hubbard model}

The two-site Hubbard model is described by the following Hamiltonian
\begin{equation}  \label{Eq3.1}
H=\sum_{ij}\left(t_{ij}-\delta_{ij}\,\mu\right)c^\dagger(i)\,c(j) +
U\sum_i n_\uparrow(i)\,n_\downarrow(i)
\end{equation}
where the summation range only over two sites at distance $a$ from
each other and the rest of notation is standard \cite{Mancini_04}.
The hopping matrix $t_{ij}$ is defined by
\begin{equation}  \label{Eq3.2}
t_{ij} = -2t\,\alpha_{ij} \;\; \;\; \;\; \alpha_{ij} = \frac12\sum_k
e^{ \mathrm{i}\,k(i-j)}\,\alpha(k)
\end{equation}
where $\alpha(k)=\cos(ka)$ and $k=0$, $\pi/a$.

We now proceed to study the system by means of the equation of
motion approach and the Green's function formalism \cite{Avella_01}.
A complete set of fermionic eigenoperators of $H$ is the following
one
\begin{equation}  \label{Eq3.4}
\psi (i)=\left(
\begin{array}{l}
\xi (i) \\
\eta (i) \\
\xi _s(i) \\
\eta _s(i)
\end{array}
\right)
\end{equation}
where
\begin{subequations}
\begin{align}  \label{Eq3.5}
&\xi(i)=\left[1-n(i)\right]c(i) \\
&\eta(i)=n(i)\,c(i) \\
&\xi _s(i)=\frac12 \sigma ^\mu\,n_\mu (i)\,\xi ^\alpha (i)+\xi
(i)\,\eta
^{\dagger\alpha}(i)\,\eta (i) \\
&\eta _s(i)=\frac12 \sigma ^\mu\,n_\mu (i)\,\eta ^\alpha (i)+\xi
(i)\,\xi^{\dagger\alpha} (i)\,\eta (i)
\end{align}
We define $\psi^\alpha(i)=\sum_j \alpha_{ij}\,\psi(j)$ and use the
spinorial notation for the field operators. $n_\mu
(i)=c^\dagger(i)\,\sigma_\mu\,c(i)$ is the charge ($\mu=0$) and spin
($\mu=1,\,2,\,3$) operator; greek (e.g., $\mu$, $\nu$) and latin
(e.g., $a$, $b$, $k$) indices take integer values from $0$ to $3$
and from $1$ to $3$, respectively; sum over repeated indices, if not
explicitly otherwise stated, is understood;
$\sigma_\mu=(1,\vec{\sigma})$ and $ \sigma^\mu=(-1,\vec{\sigma})$;
$\vec{\sigma}$ are the Pauli matrices. In momentum space the field
$\psi(i)$ satisfies the equation of motion
\end{subequations}
\begin{equation}  \label{Eq3.6}
\mathrm{i}\frac{\partial}{\partial t}\psi(k,t)=\varepsilon (k)\,\psi
(k,t)
\end{equation}
where the energy matrix $\varepsilon (k)$ has the expression
\begin{equation}  \label{Eq3.7}
\varepsilon(k)=\left(
\begin{array}{cccc}
-\mu -2t\,\alpha(k) & -2t\,\alpha(k) & -2t & -2t \\
0 & U-\mu & 2t & 2t \\
0 & 4t & -\mu +2t\,\alpha(k) & 4t\,\alpha(k) \\
0 & 2t & 2t\,\alpha(k) & U-\mu
\end{array}
\right)
\end{equation}

Straightforward calculations, according to the scheme traced in
\cite{Avella_01}, show that two correlators
\begin{align}  \label{Eq3.8}
&\Delta = \left\langle \xi^\alpha(i)\,\xi^\dagger(i)\right\rangle -
\left\langle \eta^\alpha(i)\,\eta^\dagger(i)\right\rangle \\
&p = \frac14 \left\langle n_\mu ^\alpha (i)\,n_\mu (i)\right\rangle
-\left\langle c_\uparrow (i)\,c_\downarrow (i)\left[c_\downarrow
^\dagger (i)\,c_\uparrow ^\dagger (i)\right]^\alpha\right\rangle
\end{align}
appear in the normalization matrix $I(\mathbf{k}) = \mathcal{F}
\left\langle \left\{\psi(\mathbf{i},t),\,
\psi^\dag(\mathbf{j},t)\right\}\right\rangle $. Then, the Green's
functions depend on three parameters: $\mu$, $\Delta$ and $p$. The
correlator $\Delta$ can be expressed in terms of the fermionic
correlation function $C(i,j)=\left\langle
\psi(i)\,\psi^\dagger(j)\right\rangle $; the chemical potential
$\mu$ can be related to the particle density by means of the
relation $n=2\left[1-C_{11}(i,i)-C_{22}(i,i)\right]$. The parameter
$p$ cannot be calculated in the fermionic sector; it is expressed in
terms of correlation functions of the bosonic fields $n_\mu (i)$ and
$c_\uparrow (i)\,c_\downarrow (i)$. According to this, the
determination of the fermionic Green's functions requires the
parallel study of bosonic Green's functions.

After quite cumbersome calculations, it is possible to see
\cite{Avella_01} that a complete set of bosonic eigenoperators of
$H$ in the spin-charge channel is given by
\begin{equation}  \label{Eq3.9}
B^{(\mu )}(i)=\left(
\begin{array}{l}
B_1^{(\mu )}(i) \\
\vdots \\
B_6^{(\mu )}(i)
\end{array}
\right)
\end{equation}
where
\begin{align}  \label{Eq3.10}
&B_1^{(\mu )}(i)=c^\dagger (i)\,\sigma _\mu\,c(i) \\
&B_2^{(\mu )}(i)=c^\dagger (i)\,\sigma _\mu\,c^\alpha
(i)-c^{\dagger\alpha}
(i)\,\sigma _\mu\,c(i) \\
&B_3^{(\mu )}(i)=d_\mu (i)-d_\mu ^\alpha (i)+d_\mu ^\dagger
(i)-d_\mu
^{\dagger\alpha} (i) \\
&B_4^{(\mu )}(i)=d_\mu (i)-d_\mu ^\alpha (i)-d_\mu ^\dagger
(i)+d_\mu
^{\dagger\alpha} (i) \\
&B_5^{(\mu )}(i)=f_\mu (i)-f_\mu ^\alpha (i)-f_\mu ^\dagger
(i)+f_\mu
^{\dagger\alpha} (i) \\
&B_6^{(\mu)}(i)=f_\mu(i)-f_\mu^\alpha(i)+f_\mu^\dagger(i)-f_\mu^{\dagger
\alpha}(i)
\end{align}
with the definitions:
\begin{align}  \label{Eq3.11}
d_\mu (i)&=\xi ^\dagger (i)\,\sigma _\mu\,\eta ^\alpha (i) \\
f_0(i)&=-\eta ^\dagger (i)\,\eta (i)-d^\dagger (i)\,d^\alpha (i)+\eta ^\dagger (i)\,\eta (i)\,\xi ^{\dagger\alpha} (i)\,\xi ^\alpha (i) \\
f_a(i)&=\xi ^\dagger (i)\,\xi (i)\,n_a^\alpha (i)-\frac12 \mathrm{i}
\,\epsilon_{abc}\,n_b(i)\,n_c^\alpha (i)
\end{align}
The field $B^{(\mu )}(i)$ satisfies the equation of motion
\begin{equation}  \label{Eq3.12}
\mathrm{i}\frac{\partial}{\partial t}B^{(\mu
)}(k,t)=\kappa(k)\,B^{(\mu )}(k,t)
\end{equation}
where the energy matrix $\kappa(k)$ has the expression
\begin{equation}  \label{Eq3.13}
\kappa (k)=\left(
\begin{array}{cccccc}
0 & -2t & 0 & 0 & 0 & 0 \\
-4t\left[1-\alpha(k)\right] & 0 & U & 0 & 0 & 0 \\
0 & 0 & 0 & U & 2t & 0 \\
0 & 0 & U & 0 & 0 & 2t \\
0 & 0 & 8t & 0 & 0 & 0 \\
0 & 0 & 0 & 8t & 0 & 0
\end{array}
\right)
\end{equation}

The energy spectra are given by
\begin{align}  \label{Eq3.14}
&\omega_1(k) = -2t \sqrt {2\left[1-\alpha (k)\right]} \\
&\omega_2(k) = 2t \sqrt {2\left[1-\alpha (k)\right]} \\
&\omega_3(k) = -U-4J_U \\
&\omega_4(k) = -4J_U \\
&\omega_5(k) = 4J_U \\
&\omega_6(k) = U+4J_U
\end{align}
where
\begin{equation}  \label{Eq3.15}
J_U = \frac18 \left[\sqrt {U^2+64t^2}-U \right]
\end{equation}

Straightforward calculations show that the correlation function has
the expression
\begin{multline}  \label{Eq3.16}
C^{(\mu )}(i,j) = \left\langle B^{(\mu )}(i)\,B^{(\mu
)\dagger}(j)\right\rangle \\
= \frac14 \sum_k \sum_{n=1}^6 e^{\mathrm{i}\,k(i-j)-\mathrm{i}
\,\omega_n(k)(t_i-t_j)} \left[1+\tanh \frac{\beta\,\omega _n(k)}2
\right] f^{(n,\mu)}(k)
\end{multline}
where
\begin{subequations}
\label{Eq3.17}
\begin{align}
&f^{(n,\mu )}(0) = 0 \;\; \;\; \mathnormal{for} \;\; n=3,4,5,6 \\
&f^{(n,\mu )}(\pi ) = \coth \frac{\beta\, \omega _n(\pi )}2 \sigma
^{(n,\mu )}(\pi )\;\;\forall n
\end{align}
\end{subequations}
Owing to the fact that zero-energy modes appear for $n=1$, $2$ and
$k=0$ [cfr. Eq.~(\ref{Eq3.14})], $\Gamma$ appear in the correlation
functions
\begin{equation}  \label{Eq3.18}
\Gamma^{(\mu)}(0)=\frac12 \sum_{n=1}^2 f^{(n,\mu )}(0)
\end{equation}
One might think, as is often done in the literature, to fix this
constant by its ergodic value. However, this is not correct as we
are in a finite system in the grandcanonical ensemble and the
ergodicity condition does not hold. For the moment, we can state
that this constant remains undetermined.

The spectral density functions depends on a set of parameters which
come from the calculation of the normalization matrix $I^{(\mu
)}(k)=\mathcal{F} \left\langle \left[B^{(\mu )}(i,t),\,B^{(\mu
)\dagger} (j,t)\right] \right\rangle $. In particular, for the
(1,1)-component the following parameters appear:
\begin{subequations}
\label{Eq3.17b}
\begin{align}
&C_{12}^\alpha = \left\langle \eta ^\alpha (i)\,\xi ^\dagger
(i)\right\rangle
\\
&C^\alpha= \left\langle c^\alpha (i)\,c^\dagger (i)\right\rangle \\
&d= \left\langle c_\uparrow(i)\,c_\downarrow (i)\left[c_\downarrow
^\dagger
(i)\,c_\uparrow ^\dagger (i)\right]^\alpha \right\rangle \\
&\chi^\alpha _s = \left\langle \vec{n}(i)\cdot \vec{n}^\alpha
(i)\right\rangle
\end{align}
\end{subequations}
The parameters $C^\alpha$ and $C_{12}^\alpha$ are related to the
fermionic correlation function $C(i,j)=\left\langle
\psi(i)\,\psi^\dagger(j)\right\rangle $. The parameter $\chi^\alpha
_s$ can be expressed in terms of the bosonic correlation function $
C^{(\mu)}(i,j)=\left\langle
B^{(\mu)}(i)\,B^{(\mu)\dagger}(j)\right\rangle$. In order to use the
standard procedure of self-consistency, we need to calculate the
parameter $d$. For this purpose we should open both the pair channel
and a double occupancy-charge channel (i.e., we will need the static
correlation function $\left\langle
n_\uparrow(i)\,n_\downarrow(i)\,n^\alpha(i)\right\rangle$). The
corresponding calculations are reported in Ref.~\cite{Avella_01}
where is shown that these two channels do not carry any new unknown
$\Gamma$. The self-consistence scheme closes; by considering the
four channels (i.e., fermionic, spin-charge, pair and double
occupancy-charge) we can set up a system of coupled self-consistent
equations for all the parameters. However, $\Gamma^{(\mu)}(0)$ has
not been determined yet: we have not definitely fixed the
representation of the Green's functions.

In conclusion, the standard procedure of self-consistency is very
involved and is not able to give a final answer because of the
problem of fixing the $\Gamma$. We will now approach the problem by
taking a different point of view. The proper representation of the
Green's functions must satisfy the condition that all the
microscopic laws, expressed as relations among operators must hold
also at macroscopic level as relations among matrix elements. For
instance, let us consider the fermionic channel. We have seen that
there exists the parameter $p$, not explicitly related to the
fermionic propagator, that can be determined by opening other
channels. However, we know that at the end of the calculations, if
the representation is the right one, the parameter $p$ must take a
value such that the \emph{symmetries} are conserved. By imposing the
algebra constraints (\ref{Eq1.37}) and by recalling the expression
for $\Delta$ we get three equations
\begin{subequations}
\begin{align}  \label{Eq3.25}
&n=2(1-C_{11}-C_{22}) \\
&\Delta =C_{11}^\alpha -C_{22}^\alpha \\
&C_{12}=0
\end{align}
\end{subequations}

This set of coupled self-consistent equations will allow us to
completely determine the fermionic Green's functions. Calculations
show \cite{Avella_01} that this way of fixing the representation is
the right one: all the \emph{symmetry} relations are satisfied and
all the results exactly agree with those obtained by means of Exact
Diagonalization. We do not have to open the bosonic channels; the
fermionic one is self-contained.

Next, let us consider the spin-charge Green's functions. In the
spin-charge sector we have the parameters $C^\alpha$,
$C_{12}^\alpha$, $\chi^\alpha _s$, $d$ and two $\Gamma$
\begin{align}  \label{Eq3.26}
&b_0=\frac14 \sum_{i=1}^2 f_{11}^{(i,0)}(0) \\
&b_k=\frac14 \sum_{i=1}^2 f_{11}^{(i,k)}(0) \quad \quad k=1,2,3
\end{align}
Since we are in absence of an external applied magnetic field, $b_k$
takes the same values for any value of $k$.

The parameter $C^\alpha$ and $C_{12}^\alpha$ are known, since the
fermionic correlation functions have been computed. The parameters
$\chi^\alpha _s$ and $d$ can be computed by means of the equations
\begin{align}  \label{Eq3.27}
&d=\frac14 \left\langle n_\mu ^\alpha (i)\,n_\mu (i)\right\rangle -p \\
&\chi^\alpha _s = \left\langle \vec{n}(i)\cdot \vec{n}^\alpha
(i)\right\rangle
\end{align}
The $\Gamma$ are fixed by the algebra constraints
\begin{equation}  \label{Eq3.28}
C_{11}^{(\mu )}(i,i)=\left\langle n_\mu (i)\,n_\mu (i)\right\rangle
\end{equation}
By recalling (\ref{Eq3.16}) and (\ref{Eq3.17}) we have
\begin{equation}  \label{Eq3.29}
b_\mu =\left\langle n_\mu (i)\,n_\mu (i)\right\rangle
-\frac14\sum_{i=1}^6 \left[1+\coth \frac{\beta\,\omega _i(\pi
)}2\right]\sigma_{11}^{(i,\mu )}(\pi )
\end{equation}
with
\begin{equation}  \label{Eq3.30}
\left\langle n_\mu (i)\,n_\mu (i)\right\rangle =\left\{
\begin{array}{lll}
n+2D & \mathnormal{for} & \mu =0 \\
n-2D & \mathnormal{for} & \mu =1,2,3
\end{array}
\right.
\end{equation}
$D=\left\langle n_\uparrow (i)\,n_\downarrow (i)\right\rangle $ is
the double occupancy and can be calculated by means of the fermionic
correlation functions $D=n-1+C_{11}$. Eqs.~(\ref{Eq3.27}) and
(\ref{Eq3.29}) constitute a set of coupled self-consistent equations
which will determine completely the Green's function in the
spin-charge channel. Calculations show that this way of fixing the
representation is the right one: all the symmetry relations are
satisfied and all the results exactly agree with those obtained by
means of Exact Diagonalization.

\begin{figure*}[t]
\begin{center}
\includegraphics[width=7cm]{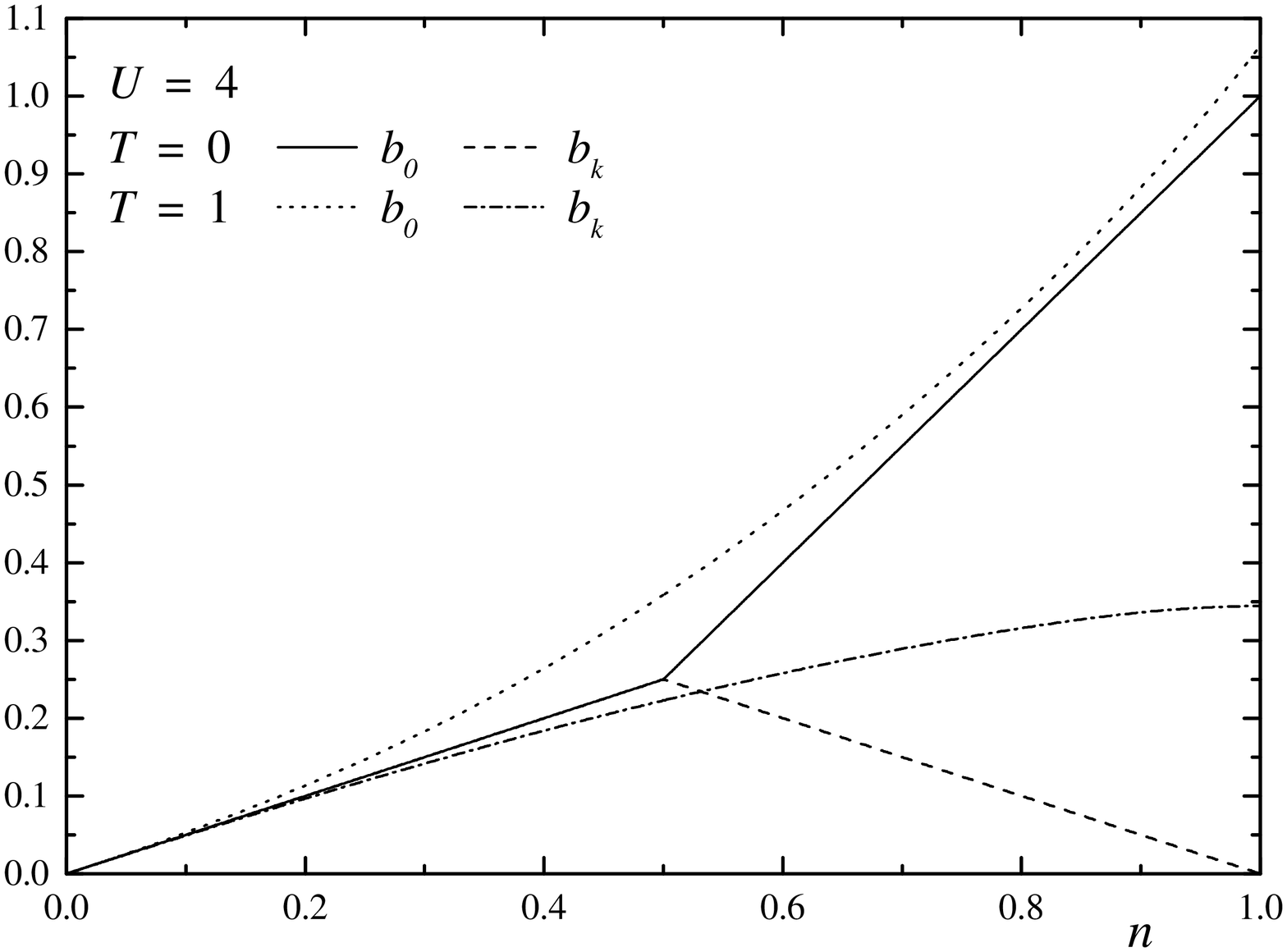}
\includegraphics[width=7cm]{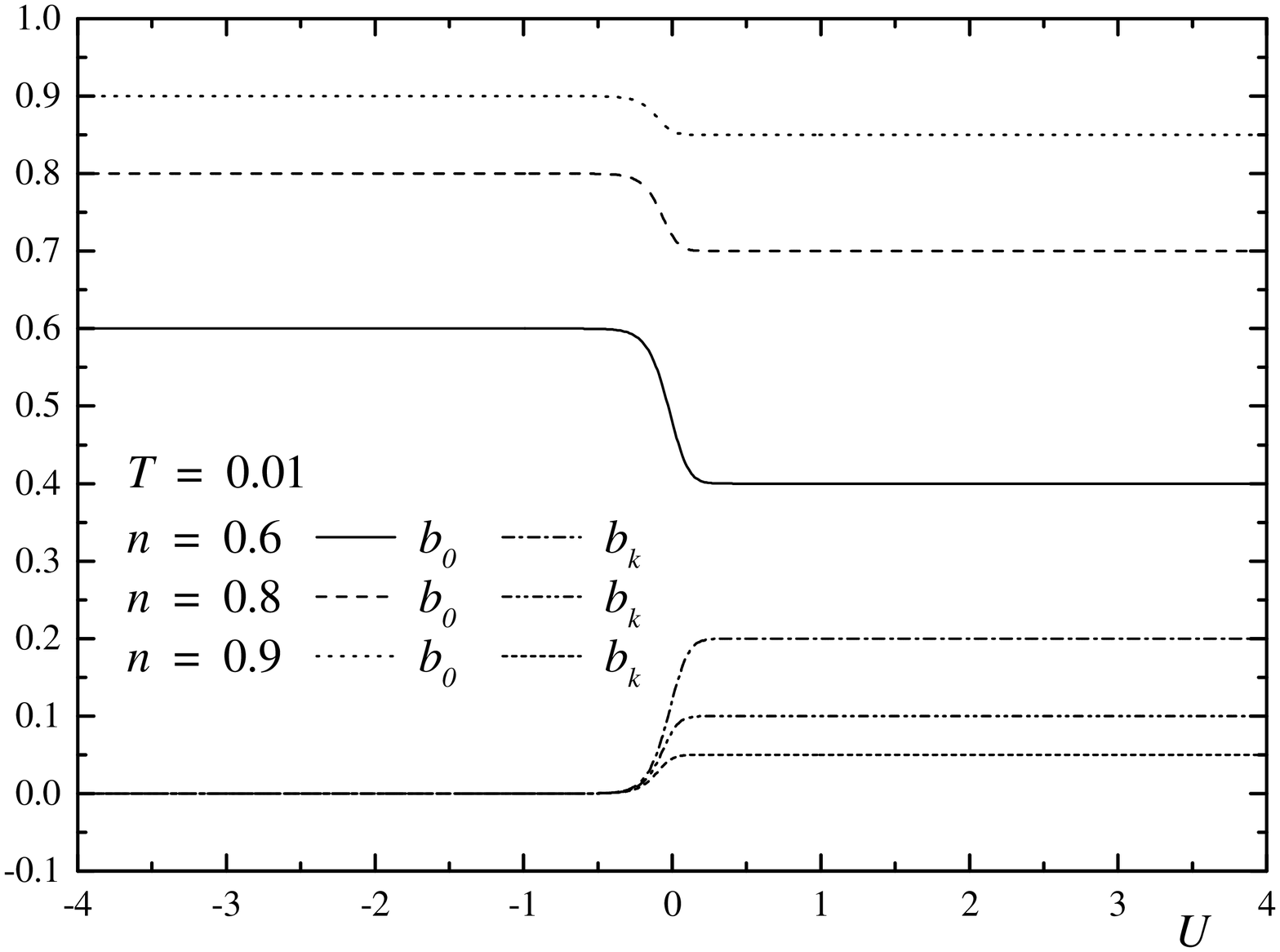}
\end{center}
\caption{(left) $b_{0}$ and $b_{k}$ are plotted as functions of $n$
for $U=4$ and $T=0$ and $1$. $U$ and $T$ are expressed in units of
$t$. (right) $b_{0}$ and $b_{k}$ are plotted as functions of $U$ for
$T=0.01$ and $n=0.6$, $0.8$, and $0.9$. $U$ and $T$ are expressed in
units of $t$.} \label{Fig_2siti}
\end{figure*}

$b_0$ and $b_k$ are plotted as functions of $n$ and $U$ in
Fig.~\ref{Fig_2siti} for various temperatures. It is worth noting
that they assume their \emph{ergodic} values (i.e. $n^2$ and $0$,
respectively) only in some regions of the parameter space: (at zero
temperature) at $n=1$ (both $b_{0}$ and $b_{k}$) and at $n=0.5$
($b_{0}$ only). In these regions, the grand-canonical ensemble is
equivalent to the microcanonical one and the underlying ergodicity
of the charge and spin dynamics emerges.

It is worth noting that $b_0$ is directly related to the
compressibility by means of the following relation \cite{Avella_01}
\begin{equation}  \label{Eq3.31}
\kappa = \frac2{k_\mathrm{B}T}\frac1{n^2}\left[b_0-n^2\right]
\end{equation}
According to this, if we erroneously set the value of $b_0$ to the
ergodic one (i.e., $n^2$) we would get a constant zero
compressibility.

\subsection{Tight-binding model}

A narrow-band Bloch system in presence of an external magnetic field
is described by the following Hamiltonian
\begin{equation}  \label{Eq4.1}
H=\sum_{\bf ij}\left(t_{\bf ij}-\mu\,\delta_{\bf ij}\right)c^\dagger
(i)\,c(j)-h\sum_{\bf i} n_3(i)
\end{equation}
where $n_3(i)$ is the third component of the spin density operator
and $h$ is the intensity of the external magnetic field. The indices
$\bf i$ and $\bf j$ run on an infinite $d$-dimensional lattice.
Straightforward calculations show that the causal Green's function
$G^{(\mu )}_C(i,j)=\left\langle
\mathcal{T}\left[n_\mu(i)\,n_\mu(j)\right] \right\rangle $ and the
correlation function $C^{(\mu )}(i,j)=\left\langle
n_\mu(i)\,n_\mu(j) \right\rangle$ of the charge-spin operator
$n_\mu(i)=c^\dagger(i)\,\sigma_\mu\,c(i)$ have the following
expressions
\begin{align}
G^{(\mu )}_C(\mathbf{k},\omega )&=-\mathrm{i}\,(2\pi)^{d+1}
a^{-d}\,\delta^{(d)}(k)\, \delta (\omega)\, \Gamma^{(\mu)} -Q^{(\mu
)}(
\mathbf{k},\omega )  \label{Eq4.2} \\
C^{(\mu )}(\mathbf{k},\omega )&=(2\pi )^{d+1}\,a^{-d}\,\delta
^{(d)}(k)\, \delta (\omega )\,\Gamma^{(\mu)}+\left[1+\tanh
\frac{\beta\, \omega}2\right] \Im \left[Q^{(\mu )}(\mathbf{ \
k},\omega )\right] \label{Eq4.3}
\end{align}
where $\delta^{(d)}(k)$ is the $d$-dimensional Dirac delta function.
$Q^{(\mu )}(\mathbf{k},\omega )$ comes from the proper fermionic
loop and is the Fourier transform of
\begin{equation}  \label{Eq4.4}
Q^{(\mu )}(i,j)=\Tr\left[\sigma _\mu\, G_C(i,j)\,\sigma _\mu\,
G_C(j,i) \right]
\end{equation}

Here $G_C(i,j)=\left\langle
\mathcal{T}\left[c(i)\,c^\dagger(j)\right] \right\rangle$ is the
causal fermionic function and has the expression
\begin{equation}  \label{Eq4.5}
G_C(\mathbf{k},\omega )=\sum_{n=1}^2 \frac{\sigma
^{(n)}}{1+e^{-\beta\,E_n( \mathbf{k})}} \left[\frac1{\omega
-E_n(\mathbf{k})+\mathrm{i}\delta} + \frac{
e^{-\beta\,E_n(\mathbf{k})}} {\omega
-E_n(\mathbf{k})-\mathrm{i}\delta} \right]
\end{equation}
with
\begin{align}
&E_1(\mathbf{k})=-\mu -2d\,t\,\alpha (\mathbf{k})-h \\
&E_2(\mathbf{k})=-\mu -2d\,t\,\alpha (\mathbf{k})+h  \label{Eq4.6} \\
&\sigma ^{(1)}=\left(
\begin{array}{cc}
1 & 0 \\
0 & 0
\end{array}
\right) \;\; \;\; \sigma ^{(2)}=\left(
\begin{array}{cc}
0 & 0 \\
0 & 1
\end{array}
\right) \label{Eq4.7}
\end{align}
where
\begin{equation}  \label{Eq6.4}
\alpha (\mathbf{k})=\frac1d \sum_{i=1}^d \cos (k_i\,a)
\end{equation}

$\Gamma^\mu$ is fixed by the algebra constraints (\ref{Eq1.37})
which requires
\begin{equation}  \label{Eq4.8}
\Gamma^{(\mu)} = \left\langle n_\mu (i)\,n_\mu (i)\right\rangle -
\frac{a^d}{(2\pi )^{d+1}} \int \! d^dk \, d\omega \left[1+\tanh
\frac{ \beta \, \omega}2\right]\Im \left[Q^{(\mu
)}(\mathbf{k},\omega )\right]
\end{equation}

The loop $Q^{(\mu )}(\mathbf{k},\omega )$ can be calculated by means
of (\ref {Eq4.5}). Calculations show
\begin{align}  \label{Eq4.9}
& \frac{a^d}{(2\pi )^{d+1}} \int \! d^dk \, d\omega \left[1+\tanh
\frac{
\beta \, \omega}2\right]\Im [Q^{(\mu )}(\mathbf{k},\omega )]  \notag \\
& =\left\langle n\right\rangle -\left\langle n_\uparrow\right\rangle
^2-\left\langle n_\downarrow\right\rangle ^2 \;\; \;\;
\mathnormal{for} \;\;
\mu=0,3 \\
& =\left\langle n\right\rangle -2\left\langle n_\uparrow
(i)\right\rangle \left\langle n_\downarrow (i)\right\rangle \;\;
\;\; \mathnormal{for} \;\; \mu=1,2
\end{align}

By recalling the algebra constraints (\ref{Eq3.30}),
Eq.~(\ref{Eq4.8}) gives for the $\Gamma$
\begin{align}  \label{Eq4.10}
& \Gamma^{(0)} = \left\langle n\right\rangle ^2 \\
& \Gamma^{(1,2)} = 0 \\
& \Gamma^{(3)} = \left\langle n_3\right\rangle ^2
\end{align}
in accordance with the ergodic nature of the spin and charge
dynamics in this system.

It is worth noting that the compressibility of this system can be
computed by means of the general formula (\ref{Eq1.36a}) that holds
in the thermodynamic limit too and gives
\begin{equation}
\kappa =\frac{1}{\left\langle n\right\rangle ^{2}}\frac{\beta
}{2}\frac{ a^{d}}{2(2\pi )^{d}}\sum_{n=1}^{2}\int
\!d^{d}k\frac{1}{C_{n}(\mathbf{k})} \label{Eq4.10a}
\end{equation}
where $C_{n}(\mathbf{k})=\cosh ^{2}\left( \frac{\beta
\,E_{n}(\mathbf{k})}{2} \right) $. We can see that an ergodic charge
dynamics can lead to a non-ergodic value of the $\Gamma$ relatively
to the total number operator, which is an integral of motion. Also
in the infinite systems the decoupling inspired by the requirement
of ergodicity cannot always be applied.

\subsection{Heisenberg chain}

We will now study \cite{Bak_02a,Plekhanoff_06} the ergodicity of the
dynamics of the operator $S^z_{\mathbf{i}}$, the $z$-component of
the spin at site $\mathbf{i}$, in the 1D anisotropic extended
Heisenberg model described by the following Hamiltonian:
\begin{equation} \label{ham}
H = -J_z \sum_\mathbf{i} S^z_\mathbf{i} S^z_{\mathbf{i}+1}+J_\perp
\sum_\mathbf{i} ( S^x_\mathbf{i} S^x_{\mathbf{i}+1} + S^y_\mathbf{i}
S^y_{\mathbf{i}+1}) +J^\prime \sum_\mathbf{i} \mathbf{S}_\mathbf{i}
\mathbf{S}_{\mathbf{i}+2},
\end{equation}
where $S^x_\mathbf{i}$, $S^y_\mathbf{i}$ and $S^z_\mathbf{i}$ are
the $x$, $y$ and $z$ components of the spin-$1/2$ at site
$\mathbf{i}$, respectively. The model~(\ref{ham}) is taken on a
linear chain with periodic boundary conditions. We take the
interaction term parameterized with $J_z$ ferromagnetic ($J_z>0$)
and the next-nearest-neighbor interaction term, which is
parameterized with $J^\prime$, isotropic. In order to frustrate
ferromagnetism, we have considered only the case with $J^\prime>0$,
that is, with an antiferromagnetic coupling between next-nearest
neighbors. According to this, only chains with even number of sites
have been studied in order to avoid topological frustration that
would be absent in the thermodynamic limit.  Since it is possible to
exactly map all results obtained for $J_\perp
> 0$ to those for $J_\perp < 0$ by means of a simple canonical
transformation, we have limited our study only to positive values of
$J_\perp$.

We have numerically diagonalized the Hamiltonian (\ref{ham}) for
chains of size $L$ ranging between $6$ and $18$ by means of Exact
Diagonalization (ED) (divide and conquer algorithm) and for chains
of size $L$ ranging between $20$ and $26$ by means of Lanczos
Diagonalization (LD). We have systematically taken into account
translational symmetry and classified the eigenstates by the average
value of $S^z=\sum_\mathbf{i} S^z_{\mathbf{i}}$, which is a
conserved quantity. Whenever we have used ED, all eigenvalues and
eigenvectors of (\ref{ham}) have been calculated up to machine
precision and, therefore, we have been able to determine the exact
dynamics of the system for all temperatures. On the contrary, when
we have used LD, we have been limited to the zero-temperature case
since only the ground state can be considered exact in LD.

In this case, we have the opportunity to exactly compute $\Gamma$ in
terms of the exact eigenvalues $E_n$ and eigenstates $|n\rangle$ of
the system. As a matter of fact, it read as \cite{Mancini_04}
\begin{equation}\label{Gammadef}
\Gamma=\frac{1}{Z}\sum_{\stackrel{n,m}{E_{n}=E_{m}}} {\rm e}^{-\beta
E_{n}} \langle n | S^z_\mathbf{i} | m \rangle \langle m |
S^z_\mathbf{i} | n \rangle
\end{equation}
As already discussed above, the dynamics of an operator (e.g.,
$S^z_\mathbf{i}$) is ergodic whenever (\ref{erg_def}) is satisfied,
or equivalently, (\ref{Gammadef}) is equal to its \emph{ergodic}
value:
\begin{equation} \label{gamma-erg}
   \Gamma^{erg}=\langle S^z_\mathbf{i} \rangle^2
   =\frac{1}{Z^2}\sum_{n,m} e^{-\beta ( E_n + E_m ) }
   \langle n | S^z_\mathbf{i} | n \rangle
   \langle m | S^z_\mathbf{i} | m \rangle.
\end{equation}
The dynamics of a finite system is hardly ergodic, since
(\ref{Gammadef}) and (\ref{gamma-erg}) unlikely coincide. In the
thermodynamic limit, the sums in (\ref{Gammadef}) and
(\ref{gamma-erg}) become series and no conclusion can be drawn
\textsl{a priori}. Since we have diagonalized the Hamiltonian
(\ref{ham}) numerically (i.e., only for finite systems) and since $L
\to \infty$ is the most interesting case, we have analyzed our
results through finite-size scaling in order to speculate on the
properties of the bulk system.

If the ground state of (\ref{ham}) is $N$-fold degenerate then, at
$T=0$, (\ref{Gammadef}) and~(\ref{gamma-erg}) read as follows:
\begin{eqnarray*} \label{t0-av}
   \Gamma^{\phantom{erg}}&=&\frac{1}{N} \sum_{n, m=1}^{N}
   |\langle n | S^z_\mathbf{i} | m \rangle |^2 \\
   \Gamma^{erg}&=&\left ( \frac{1}{N} \sum_{n}^{N}
   \langle n | S^z_\mathbf{i} | n \rangle \right )^2,
\end{eqnarray*}
respectively.

Thanks to the translational invariance enjoined by the system
$\langle S^z_\mathbf{i} \rangle$ is independent of $\mathbf{i}$ and
proportional to the $z$-component of the total spin operator average
$\langle S^z_{tot} \rangle$. It is easy to show that, even if there
is a finite magnetic moment per site in any of such $N$ degenerate
ground states, $\langle S^z_\mathbf{i} \rangle$ at $T =0$ is always
zero in absence of magnetic field. Indeed, if a ground state with
non-zero $\langle S^z_\mathbf{i} \rangle = M$ exists, also another
ground state with $\langle S^z_\mathbf{i} \rangle = -M$ exists.
Thus, at zero temperature, $\Gamma^{erg}$ is always zero and the
only quantity of interest is $\Gamma$. A finite value of this latter
implies non-ergodicity. Obviously, if $N=1$ then both values
coincide. Therefore, a non-ergodic phase corresponds to degenerate
ground states with finite magnetization.

\begin{figure}
\begin{center}
\includegraphics[width=10cm]{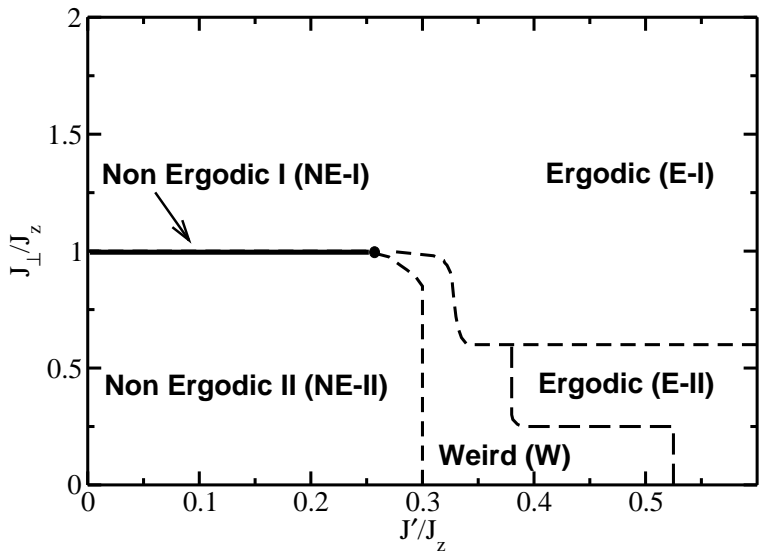}
\end{center}
\caption{Zero-temperature ergodicity phase diagram in
the $J^{\prime}-J_{\perp}$ plane. Due to the symmetry of the
Hamiltonian only the upper half is shown (see in the text). Only two
ergodic phases (E-I and E-II) have been found in the reported
parameter space. The others are either non-ergodic (NE-I and NE-II)
or impossible to conclusively analyze (W). The latter phase might
shrink to a transition line in the bulk limit.} \label{fig1}
\end{figure}

In the studied range of coupling constants (see Fig.~\ref{fig1}) we
have found two non-ergodic phases (NE-I and NE-II), two ergodic ones
(E-I and E-II) and a \emph{weird} phase (W). Our computational
facilities limit the range of chain sizes that we can analyze such
that we could not establish, by means of finite-size scaling,
whether the \emph{weird} phase (W) is ergodic or not. In the
non-ergodic phases (NE-I and NE-II), we were able not only to
perform the finite-size scaling, but also to write down an analytic
expression for $\Gamma$ as a function of the chain size $L$. The
\emph{weird} phase (W) has exhibited a strong dependence of the
ground state upon the particular values of the couplings. On the
contrary, the other phases exhibit ground states that are
independent of the particular values of the coupling constants.

\begin{figure*}[t]
\begin{center}
\includegraphics[width=7cm]{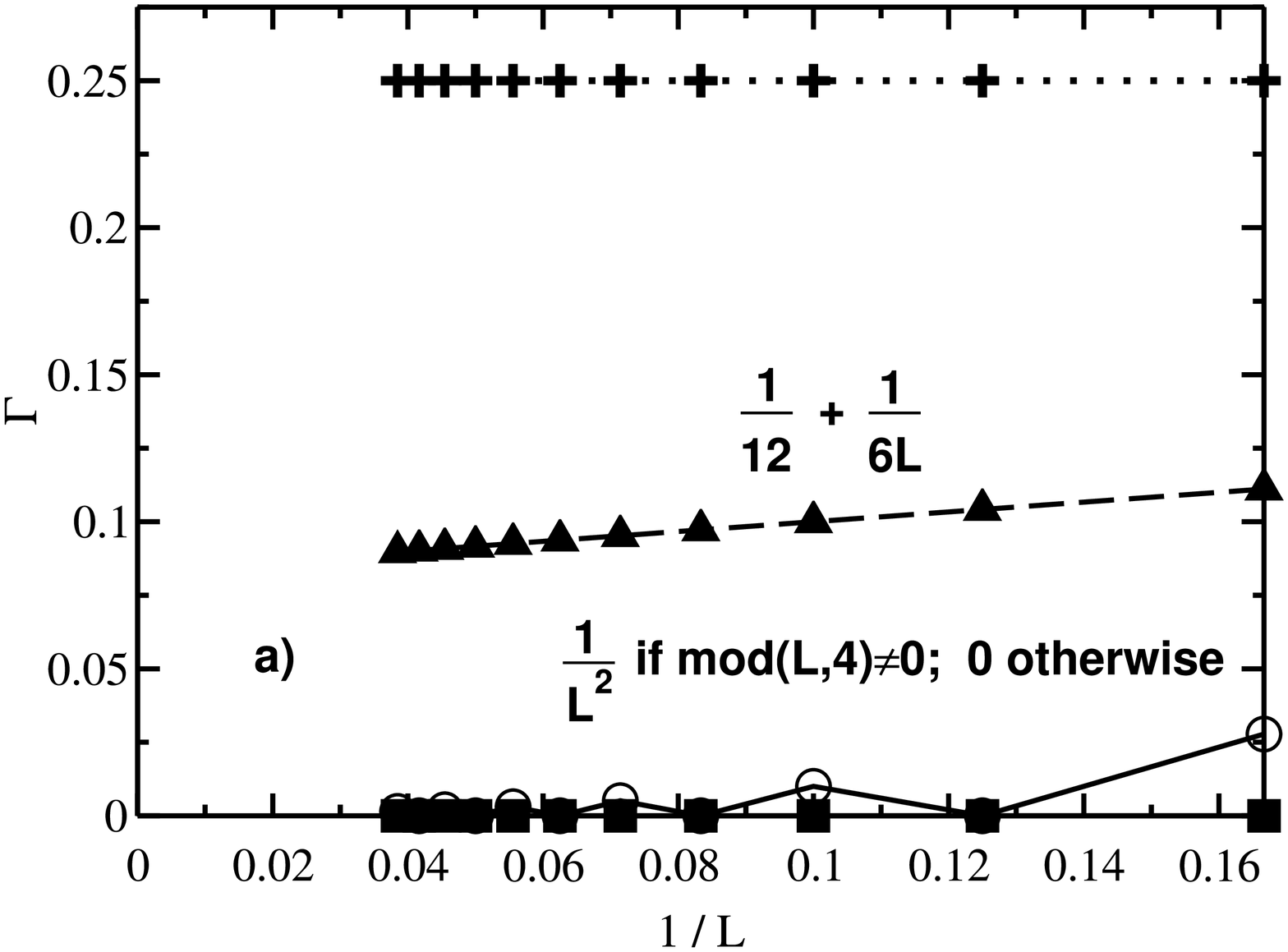}
\includegraphics[width=7cm]{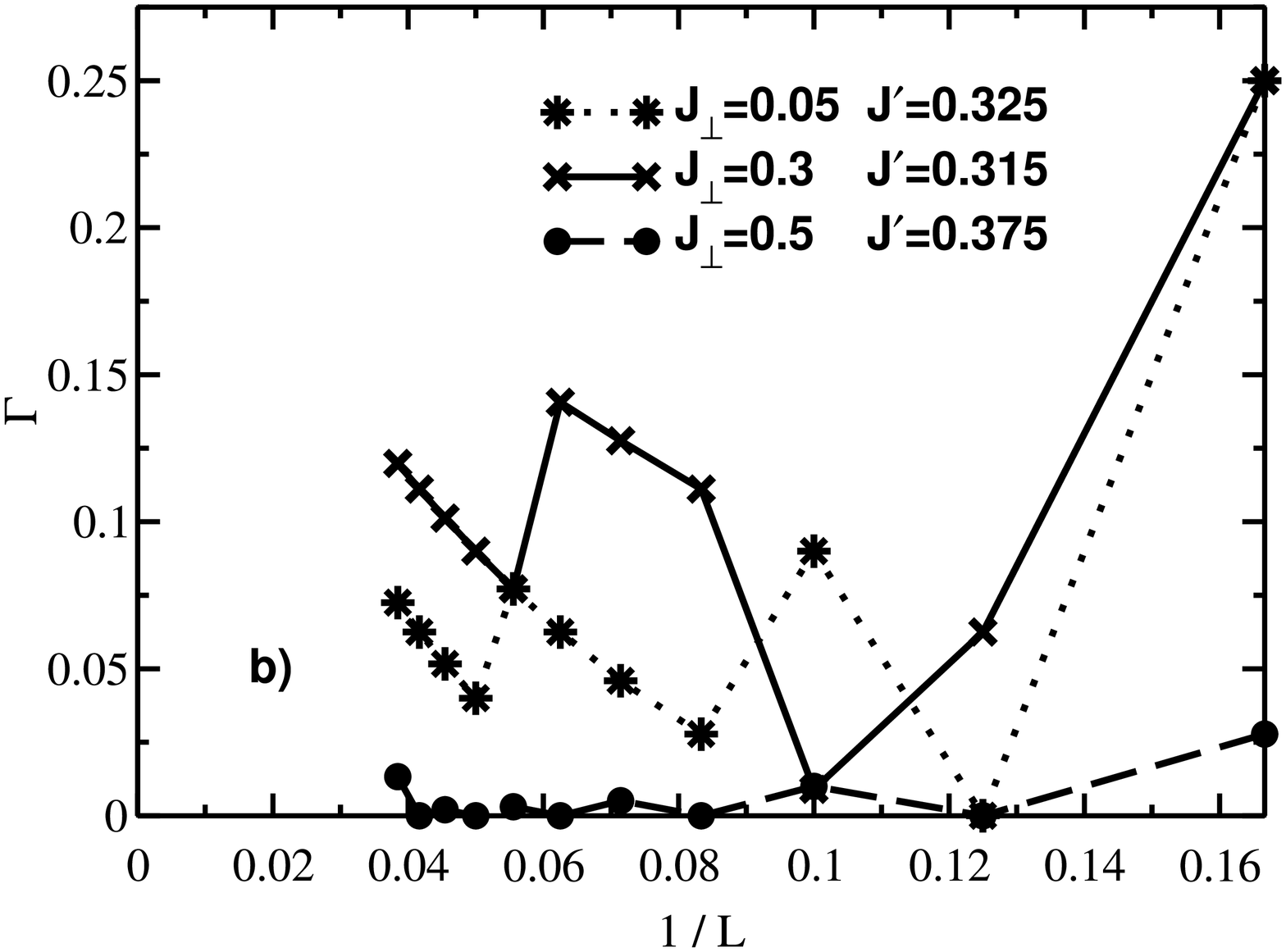}
\end{center}
\caption{\label{fig2} Finite-size scaling in the case of $T=0$ for
different points in the phase diagram of Fig.~\ref{fig1}. Symbols on
panel a): $+$ corresponds to (NE-II), $\blacktriangle$ corresponds
to (NE-I), $\blacksquare$ corresponds to (E-I) and $\bigcirc$ to
(E-II) regions of Fig.~\ref{fig1}, respectively. On panel b)
different examples from (W) region are shown. Hamiltonian couplings
are shown in the legend. All energies are expressed in units of
$J_z$.}
\end{figure*}

In the standard Heisenberg model ($J^\prime=0$ and $J_\perp=J_z$) at
$T=0$ the dynamics is non-ergodic for ferromagnetic coupling
($J_\perp=J_z<0$) as the system has a $L+1$ degenerate ground state
\begin{equation} \label{heis_ne}
\Gamma = \frac{1}{12} + \frac{1}{6L}.
\end{equation}
It is clear from (\ref{heis_ne}) that $\Gamma$ remains non-ergodic
also in the thermodynamic limit. This point ($J^\prime=0$ and
$J_\perp=J_z$) becomes a line in our phase diagram and is denoted as
NE-I (see Fig.~\ref{fig1}). In fact, the next-nearest-neighbor
interaction $J^\prime$ may frustrate ($J^\prime>0$) or favor
($J^\prime<0$) the ferromagnetism. In the latter case, the ground
state remains unchanged for any value of $J^\prime<0$. Therefore, we
expect the line denoting the phase NE-I to extend also to negative
$J^\prime$. If, on the contrary, $J^\prime$ is positive and large
enough to frustrate the system in such a way that the ground state
loses its ferromagnetic character, the ergodicity is restored. This
occurs at a finite critical $J^\prime \sim 0.25 J_z$. For values of
$J^\prime$ larger than the critical one, we find a non-degenerate
ground state with $\langle S^z_{tot} \rangle=0$.

If $J_\perp \neq J_z$ the rotational invariance is broken so that
states with the same $\langle {\bf S}_{tot}^2 \rangle$, but
different $\langle S_{tot}^z \rangle$, are not degenerate anymore.
In the non-ergodic region (NE-II) of the phase diagram (see
Fig.~\ref{fig1}), the ground state is just doubly degenerate (not
$L+1$ degenerate as in (NE-I)): one ground state corresponds to a
configuration with all spins \emph{up} and the other to a
configuration with all spins \emph{down}. Hence, the value of
$\Gamma$ in this phase is $1/4$ and does not depend neither on the
Hamiltonian couplings nor on the number of sites in the chain. It is
clear that also this phase extends to negative values of $J^\prime$.
This kind of ground state stands the frustration introduced by
next-nearest-neighbor interaction up to $J^\prime \sim 0.3J_z$ (see
Fig.~\ref{fig1}).

The ergodic region (E-I) of the phase diagram (see Fig.~\ref{fig1})
has $\Gamma=0$ for all sizes of the system and values of the
couplings: the unique ground state belongs to the sector with
$\langle S_{tot}^z \rangle = 0$. On the contrary, the other ergodic
phase (E-II) has non-zero values of $\Gamma$ for values of $L$ not
multiples of four. The ground state in this phase has average total
spin equal to one and, therefore, $\Gamma=1/L^2$. We obviously
conclude that (E-II) phase is ergodic in the thermodynamic limit.

The values of $\Gamma$ in these four phases (NE-I, NE-II, E-I and
E-II) exhibit perfect finite-size scaling as shown in
Fig.~\ref{fig2}a). This has allowed us to make definite statements
also in the thermodynamic limit.

The \emph{weird} phase (W) (see Fig.~\ref{fig1}) is characterized by
a quite strong size dependence, as shown in Fig.~\ref{fig2}b) where
a tentative finite-size scaling of $\Gamma$ in the different points
of the phase is presented. This region manifests a diverging
finite-size scaling within the range of sizes we were able to
handle. In this case, the behavior of $\Gamma$ as a function of $L$
strongly depends on the particular choice of the Hamiltonian
couplings and is highly non monotonous when increasing $L$,
according to the strong dependence on $L$ of $\langle S_{tot}^z
\rangle$ in the ground state. In this critical region the
eigenvalues of (\ref{ham}) present many level crossings, which means
that the maximum value of $L$ we were able to reach ($L_{max}=26$)
is not large enough to perform a sensible finite-size scaling
analysis. However, we expect that this phase becomes ergodic in the
thermodynamic limit, although still different from the ergodic
phases E-I and E-II.

We can summarize our findings in the thermodynamic limit at zero
temperature as follows:
\begin{equation}\label{gcase}
\Gamma=
\begin{cases}
\frac{1}{12} & \text{if $J_\perp= \pm J_z$ and $J^\prime \lesssim 0.25 J_z$} \\
\frac{1}{4} & \text{if $|J_\perp| < J_z$ and $J^\prime \lesssim 0.3 J_z$} \\
??? & \text{in the \emph{weird} phase (W) (see Fig.~\ref{fig1})}\\
0 & \text{otherwise}
\end{cases}
\end{equation}

\section{Conclusions}

In conclusion, we have analyzed the issue of ergodicity, after a
brief historical overview, within the Green's function formalism by
means of the equations of motion approach. We have individuated the
primary source of non-ergodic dynamics for a generic operator in the
appearance of zero-frequency anomaly in its correlation functions
and given a recipe to compute the unknown quantities characterizing
such a behavior within the Composite Operator Method. Finally, we
have presented examples of non-trivial strongly correlated systems
where it is possible to examine a non-ergodic behavior: two-site
Hubbard model, tight-binding model, Heisenberg chain.


\label{last@page}
\end{document}